\documentclass[preprint]{aastex63}

\usepackage{graphicx} 
\usepackage{colortbl,booktabs}
\usepackage{indentfirst}
\usepackage{gensymb} 
\usepackage{natbib}
\usepackage{lineno}

\shorttitle{Two Homologous Magnetic Flux Ropes}
\shortauthors{De-Chao Song et al.}

\graphicspath{{./}{figures/}}

\begin{document}
\title{Multi-wavelength and Dual-perspective Observations of Eruption and Untwisting of Two Homologous Magnetic Flux Ropes}

\author[0000-0003-0057-6766]{De-Chao Song}
\affiliation{Key Laboratory of Dark Matter and Space Astronomy, Purple Mountain Observatory, CAS, Nanjing 210023, People's Republic of China}
\affiliation{School of Astronomy and Space Science, University of Science and Technology of China, Hefei 230026, People's Republic of China}
\author[0000-0002-8258-4892]{Y. Li}
\affiliation{Key Laboratory of Dark Matter and Space Astronomy, Purple Mountain Observatory, CAS, Nanjing 210023, People's Republic of China}
\affiliation{School of Astronomy and Space Science, University of Science and Technology of China, Hefei 230026, People's Republic of China}
\author{Y. Su}
\affiliation{Key Laboratory of Dark Matter and Space Astronomy, Purple Mountain Observatory, CAS, Nanjing 210023, People's Republic of China}
\affiliation{School of Astronomy and Space Science, University of Science and Technology of China, Hefei 230026, People's Republic of China}
\author[0000-0002-4978-4972]{M. D. Ding}
\affiliation{School of Astronomy and Space Science, Nanjing University, Nanjing 210023, People's Republic of China}
\author{W. Q. Gan}
\affiliation{Key Laboratory of Dark Matter and Space Astronomy, Purple Mountain Observatory, CAS, Nanjing 210023, People's Republic of China}
\affiliation{School of Astronomy and Space Science, University of Science and Technology of China, Hefei 230026, People's Republic of China}

\correspondingauthor{Y. Li}
\email{yingli@pmo.ac.cn}

\begin{abstract}
In this paper, we present a detailed morphological, kinematic, and thermal analysis of two homologous magnetic flux ropes (MFRs) from NOAA 11515 on 2012 July 8--9. The study is based on multi-wavelength and dual-perspective imaging observations from the Solar Dynamics Observatory and the Solar Terrestrial Relations Observatory Ahead spacecraft, which can well reveal the structure and evolution of the two MFRs. We find that both of the MFRs show up in multiple passbands and their emissions mainly consist of a cold component peaking at a temperature  of $\sim$0.4--0.6 MK and a hot component peaking at $\sim$7--8 MK. The two MFRs exhibit erupting, expanding, and untwisting motions that manifest distinctive features from two different viewpoints. Their evolution can be divided into two stages, a fast-eruption stage with speeds of about 105--125 km s$^{-1}$ for MFR-1 and 50--65 km s$^{-1}$ for MFR-2 and a slow-expansion (or untwisting) stage with speeds of about 10--35 km s$^{-1}$ for MFR-1 and 10--30 km s$^{-1}$ for MFR-2 in the plane of sky. We also find that during the two-stage evolution, the high temperature features mainly appear in the interface region between MFRs and ambient magnetic structures and also in the center of MFRs, which suggests that some heating processes take place in such places like magnetic reconnection and plasma compression. These observational results indicate that the eruption and untwisting processes of MFRs are coupled with the heating process, among which an energy conversion exists.
\end{abstract}


\section{Introduction}
\label{intro}

Helical magnetic flux ropes (MFRs) are conventionally related to filaments (or prominences), which are one kind of the fundamental structures in the standard solar flare (or CSHKP) model \citep{Carmichael1964,Sturrock1966,Hirayama1974,KP1976}. Erupting MFRs and filaments are sometimes observed to experience a rotation as they arise \citep [e.g.,][]{Vrsnak1980,Kuro1987,Zhou2006,Green2007,Muglach2009,Bemporad2011,Su2013,Yan2013}. Some tornado-like rotational movements are also detected in filaments with high-resolution observations \citep [e.g.,][]{Li2012,Su2012,Wedemeyer2013,Panesar2013,Su2014}. These rotations are usually interpreted as a supply of twist into MFRs or a transformation of twist into writhe of MFRs. Note that twist is an inherent property of MFRs, which is strongly linked to the magnetic free energy and filament eruptions \citep[e.g.,][]{Prior2020ar,MacTaggart2020}.

In a manner of speaking, one of the key processes  determining the stability and eruption behaviors of MFRs is untwisting or unwinding. The untwisting motion together with the mass flow that reveals a twisted structure was commonly observed in many events as reported by \cite{McCauley2015}. This process is often closely associated with eruptions of filaments and flares which are usually accompanied by coronal mass ejections (CMEs) \citep [e.g.,][] {Sakurai1976,Hood1992,Torok2005,Kliem2006,Dere2009}. There are also some reports about the untwisting process of erupting MFRs that are unrelated to CMEs \citep [e.g.,][] {Ji2003,Alexander2006,Yan2020a}. In a rough classification, the untwisting motions fall into symmetric \citep [e.g.,][] {Martin2003} and asymmetric \citep [e.g.,][] {Tripathi2006,Bi2013} types depending on whether they erupt at the top or footpoint of the filaments \citep [e.g.,][] {McCauley2015}. These investigations indicate that the untwisting of MFRs plays a pivotal role in solar eruptions. However, how the MFRs are untwisted is still under discussion and this needs more observations to explore.

Some researchers have investigated the untwisting process of MFRs associated with solar jets \citep{Patsourakos2008,Nistic2009,Chen2012,Curdt2012,Morton2012,Shen2012,Lee2013,Liu2014,Zhang2014,Zhu2017}. \cite{Chen2017} also reported rapid rotating and spinning of magnetic field structures, which  were triggered by an interaction between EUV jets and filaments. Moreover, 3D magnetohydrodynamic (MHD) models of jets \citep[e.g.,][]{Pariat2009,Pariat2010,Rachmeler2010,Pariat2015,Karpen2017} likewise produce helical and untwisting structures that are similar to the observed ones. In recent high-resolution observations, the untwisting of MFRs in filament eruptions has been reported \citep[e.g.,][] {Kumar2012,Li2013,Cheng2016,Duchlev2016,Xu2017,Chen2019,Yan2020b}. Some researchers also utilized the multi-viewpoint observations from the Solar Dynamics Observatory \citep[SDO;][]{Pesnell2012} and/or the Solar Terrestrial Relations Observatory \citep[STEREO;][]{Howard2008,Kaiser2008} to investigate the continuous evolution of MFRs \citep[e.g.,][]{Bemporad2011,Thompson2011,Joshi2011,Su2013,Zhou2017,Wang2019}. Nevertheless, they have rarely focused on the untwisting process along with the thermal property of MFRs .

The thermal property of MFRs during the eruption can be studied via the differential emission measure (DEM) method \citep[e.g.,][]{Golub2004,Weber2004,HK2012,Cheung2015dem,Su2018} which is a powerful tool for extracting plasma parameters such as emission measure (EM), DEM- or EM-weighted temperature, and electron density. For example, \cite{cheng2012} estimated the temperature and density of the multi-structure components of CMEs (or MFRs) using the DEM method. Their results show that the core regions of CME are dramatically heated, presumably via magnetic reconnection, and the DEM-weighted temperature of the MFR centroid increases from $\sim$8.0 MK to $\sim$10.0 MK during the eruption. \cite{Krucker2014} deduced the DEM of the above-the-loop-top source that shows both cold and hot components. They claimed that the hot component is most likely connected with an M7.7 flare. 

In this paper, we investigate two homologous filament eruptions associated with flares as well as two successive MFRs with multi-wavelength and dual-perspective imaging observations from SDO and STEREO-A. The two MFRs exhibit a similar morphological evolution in the field of view (FOV) of STEREO-A but a different one in the FOV of SDO. These comprehensive observations will give us a better understanding of the morphological evolution as well as kinematic and thermal properties of the two MFRs in their erupting processes. This paper is organized as follows. In Section \ref{data}, we describe the observational data and method. The analysis and results are shown in Section \ref{res}. Sections \ref{sum} gives the summary and discussions.

\section{Observational Data and Method}
\label{data}

The two homologous MFRs from NOAA active region 11515 \citep{Louis2014,Wangya2018} and their related eruptions of filaments and flares were simultaneously observed by SDO and STEREO-A that had a separation angle of $\sim$119.7\degree\ during 2012 July 8--9. The imaging data from the Atmospheric Imaging Assembly \citep[AIA;][]{Lemen2012} on board SDO and the Extreme Ultraviolet Imager \citep[EUVI;][]{Wuelser2004} on board STEREO-A are used in this study. AIA provides full-disk EUV and UV images in multiple channels with a high spatial resolution of 0.6\arcsec\ pixel$^{-1}$ and a high temporal resolution of 12 s or 24 s. The multiple-wavelength images from AIA are sensitive to temperatures ranging from 0.05--20 MK. Here we mainly use the images from one UV channel centered at 1600 \AA\ (\ion{C}{4}, $\sim$0.1 MK) and seven EUV channels centered at 94 \AA\ (\ion{Fe}{18}, $\sim$7 MK), 131 \AA\ (\ion{Fe}{8}, $\sim$0.4 MK; \ion{Fe}{21}, $\sim$11 MK), 171 \AA\ (\ion{Fe}{9}, $\sim$0.6 MK), 193 \AA\ (\ion{Fe}{12}, $\sim$1.3 MK; \ion{Fe}{24}, $\sim$20 MK), 211 \AA\ (\ion{Fe}{14}, $\sim$2 MK), 304 \AA\ (\ion{He}{2}, $\sim$0.05 MK), and 335 \AA\ (\ion{Fe}{16}, $\sim$2.5 MK). The 1.5 level data of AIA are analyzed. EUVI obtains full-disk EUV images in four channels (sensitive to plasmas at 0.1--20 MK) with a spatial resolution of $\sim$1.6\arcsec\ and a temporal resolution of 3--6 minutes. Here we only use the 304 \AA\ (\ion{He}{2}, $\sim$0.05 MK) and 195 \AA\ (\ion{Fe}{12}, $\sim$1.3 MK; \ion{Fe}{24}, $\sim$20 MK) images from EUVI to show the evolution of two eruption events.

Using the six AIA EUV channels (excluding the one at 304 \AA), we employ the DEM method as introduced in \cite{Su2018} to diagnose the thermal property of the two MFRs. This method is developed based on the one from \cite{Cheung2015dem} and can well constrain the DEMs at high temperatures by using AIA data only. We have binned the imaging data by 2 $\times$ 2 pixels to improve the signal-to-noise ratio when constructing the EM and EM-weighted temperature maps. 
For the temperature $T$ (in units of K), we set a bin of 0.05 in logarithmic scale. The EM (in units of cm$^{-5}$) is defined by
\begin{equation}
EM= \int DEM(T)dT= \int n_{e}n_{H}dl\propto n{_{e}}^{2},
\end{equation}
where $n_{e}$ and $n_{H}$ are the number densities of electron and hydrogen, respectively, and $l$ is the optical depth along the line-of-sight (LOS).
The DEM and EM-weighted temperature are defined by
\begin{equation}
DEM_{i} =\frac{EM_{i}}{\Delta T_{i}}=\frac{EM_{i}}{T_{i}ln10\Delta logT}
\end{equation}
and
\begin{equation}
\bar{T}=\frac{\sum (EM_{i}T_{i})}{\sum EM_{i}},
\end{equation}
where $i$ is the $i$th bin of log $T$ (also see the formulas in  \citealt{sun2014,Su2018,Xue2020}).

\section{Analysis and Results}
\label{res}

\subsection{Overview of the Two Homologous Events}
\label{overview}

\begin{figure}
\centering
\includegraphics[width=0.75\textwidth]{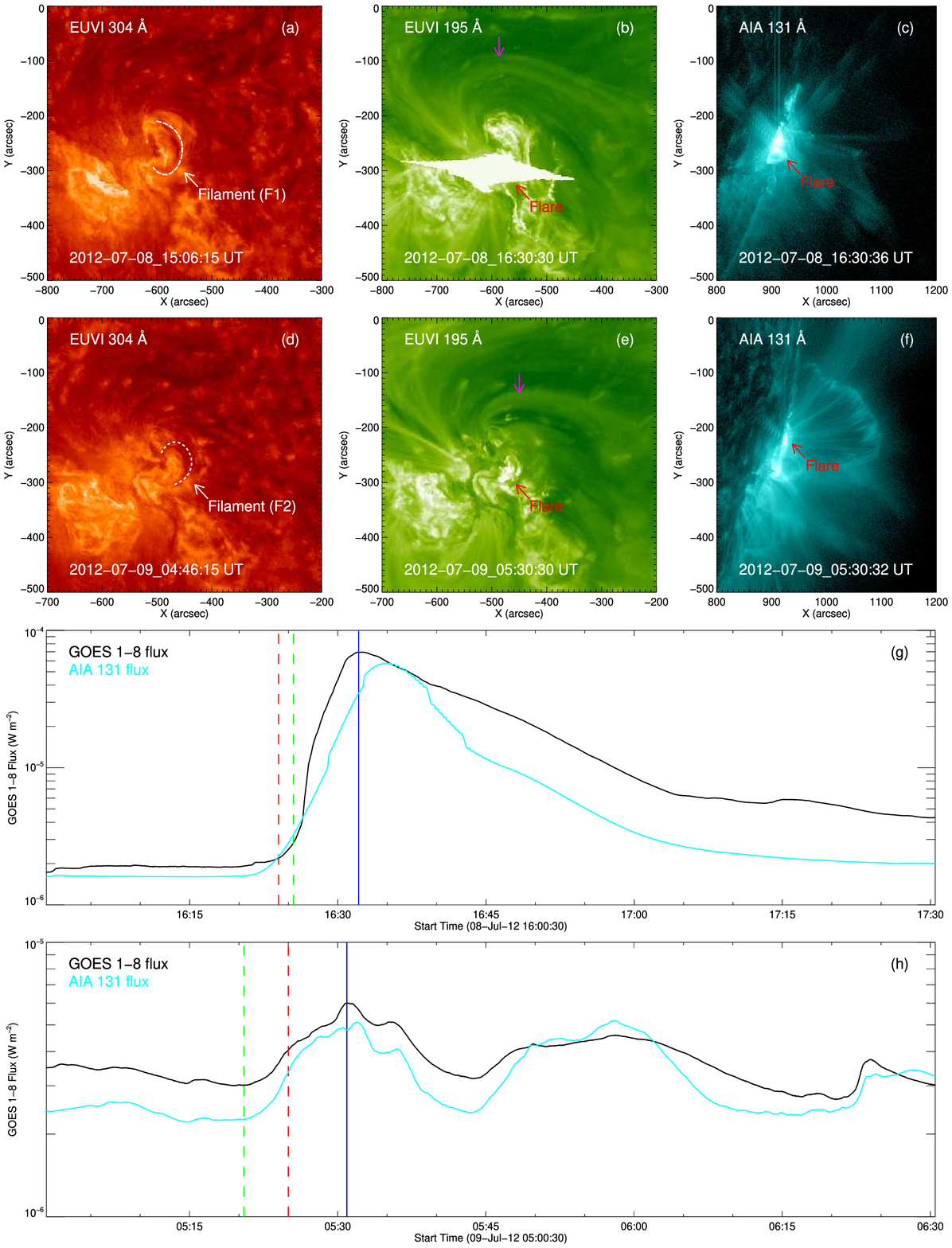}
\\[0mm]
\caption{An overview of the two eruption events. Panels (a)--(f): STEREO-A/EUVI and SDO/AIA images showing the two eruptions from the face view and side view, respectively. The white curves and arrows indicate the two filaments. The red arrows mark the associated flares around the GOES peak time. The magenta arrows denote the wrapping magnetic fields as seen in EUVI images. Panels (g) and (h): GOES 1--8 \AA\ SXR light curves (black) showing the flares. The cyan curves represent the integrated fluxes at 131 \AA\ over the regions as shown in panels (c) and (f) respectively. The vertical blue lines indicate the GOES peak times of the two flares. The vertical green and red lines denote the moments at which the flares can be seen in EUVI and AIA images, respectively.
The images shown in panels (b) and (e) are available as an animation which depicts a temporal evolution of the two eruption events, including the lightening of flares and eruption of MFRs in STEREO-A/EUVI 195 \AA\ waveband from the face view. This online animation runs from 16:00 UT to 17:30 UT on July 8 for MFR-1 and from 05:00 UT to 06:30 UT on July 9 for MFR-2 with a cadence of 5 min. The real-time duration of this animation is about 4 s.
}
\label{f1}
\end{figure}

Figure \ref{f1} gives the overview of the two homologous eruption events. On 2012 July 8--9, NOAA 11515 was located near the southwest limb from the perspective of SDO (i.e., side view, Figures \ref{f1}(c) and (f)) but appeared on the solar disk in the FOV of STEREO-A (face view, Figures \ref{f1}(a), (b), (d), and (e)), which enables us to study the two events from dual perspectives. From the EUVI 304 \AA\ images (Figures \ref{f1}(a) and (d)) we can see that two filaments (referred to as F1 for the filament on July 8 and F2 for the one on July 9, indicated by the white dashed curve) show up at an early time. Tens of minutes later, the two filaments erupt (see the accompanying animation of Figure \ref{f1}), which are accompanied by two flares, an M6.9 one and a C6.0 one, respectively (as revealed from the EUVI 195 \AA\ and AIA 131 \AA\ images in Figures \ref{f1}(b), (c), (e), and (f) and also from the GOES soft X-ray light curves in Figures \ref{f1}(g) and (h)). In the meantime, two helical structures, i.e., two twisted MFRs (called MFR-1 and MFR-2 hereafter), appear and exhibit some eruption and expansion motions, as clearly shown in Figures \ref{f2} and \ref{f6}, respectively (also see the accompanying animations). The two MFRs also show untwisting motions that last for a few tens of minutes (Figures \ref{f3} and \ref{f7}).  During the erupting and untwisting processes, some of the MFR plasmas are heated into a high temperature (Figures \ref{f4}, \ref{f5}, \ref{f8}, and \ref{f9}). At a late stage, there are materials falling back along the magnetic structures. It should be noted that the two homologous MFRs appear in a similar magnetic environment such as being wrapped by some magnetic field lines with a large spiral arm (marked by the magenta arrow in Figures \ref{f1}(b) and (e)). They also have many similar behaviors, say, exhibiting helical motions and showing a close association with eruptions of filaments and flares, as shown in the EUVI images in Figure \ref{f1} and the accompanying animation. There are, however, some differences between them, which can clearly be seen in the AIA images, as described below.

\subsection{The MFR-1 on 2012 July 8}

\subsubsection{Morphological evolution}
\label{Morphological evolution}

\begin{figure}[ht]
\centering
\includegraphics[width=0.55\textwidth]{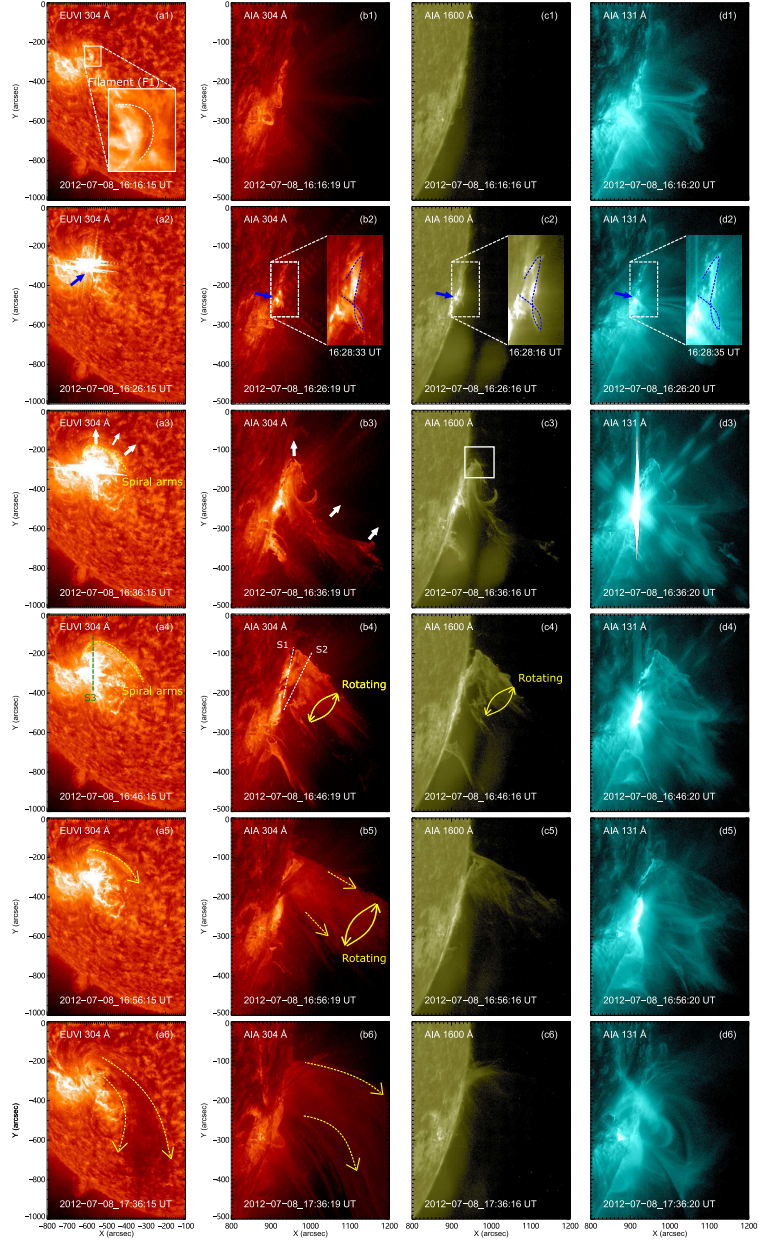}
\caption{Snapshots of the onset, development, and disintegration of the first event at EUVI 304 \AA, AIA 304, 1600, and 131 \AA\ (from left to right) from 16:16 UT to 17:36 UT on 2012 July 8. In panel (a1), the filament (F1) is enlarged and indicated by a white dashed curve. The blue arrows in panels (a2)--(d2) denote the associated flare. In panels (b2)--(d2), the reversed-$\gamma$ shape MFR-1 is enlarged and marked by a blue dashed curve. The white arrows in panels (a3) and (b3) indicate the eruption and moving direction of the materials of MFR-1. The white box in panel (c3) encloses the region that is used to show the swirling motions and material ejections of MFR-1 in Figures \ref{f3}(b) and (d). The three slices S1, S2, and S3, indicated in panels (a4) and (b4), are used to study the temporal evolution of MFR-1 as shown in Figures \ref{f3}(a), (c), and (e), respectively. The yellow dashed curves in panels (a3) and (a4) refer to some magnetic structures with a spiral arm. The yellow solid arrows in panels (b4), (c4), and (b5) indicate the rotation of MFR-1 and the yellow dashed arrows in panels (a5), (b5), (a6), and (b6) mark the ejection of the materials from MFR-1. 
The images with the same wavelength selection are available as an animation which presents a temporal evolution of MFR-1, including the onset, development, and disintegration processes, at EUVI 304 \AA\ (with a cadence of 10 min), AIA 304 (a cadence of 12 s), 1600 (a cadence of 24 s), and 131 \AA\ (a cadence of 12 s) (from left to right in the animation) from 16:16 UT to 17:46 UT on 2012 July 8. The real-time duration of this animation is about 15 s.
}
\label{f2}
\end{figure}

The first eruption event took place on 2012 July 8. Figure \ref{f2} shows some snapshots for its onset, development, and disintegration at EUVI 304 \AA\ as well as AIA 304, 1600, and 131 \AA\ from 16:16 UT to 17:36 UT. Before the eruption, there exists a reversed-C shape filament (F1), as indicated by the white dashed curve in Figure \ref{f2}(a1). At $\sim$16:24 UT, F1 erupts as displaying a reversed-$\gamma$ shape (indicated by the blue dashed curve in Figures \ref{f2}(b2)--(d2)) and brightens up implying a possible heating of its materials. In the meanwhile, an M6.9 flare (denoted by the blue arrow in Figures \ref{f2}(a2)--(d2)) occurs near F1. Afterwards, F1 rises rapidly with an inclination to the north, showing a helical structure, i.e., MFR-1, as clearly revealed in the multi-temperature AIA images (Figures \ref{f2}(b3)--(d3) and \ref{f3}(b)). At the moment, a cluster of spiral arms (outlined by the dashed curve in Figures \ref{f2}(a3) and (a4)) come into being in a face view of EUVI. In particular, there appears a clockwise swirling motion in the north part of MFR-1 from the side view of AIA (marked by a white box in Figure \ref{f2}(c3)), which is clearly shown in Figure \ref{f3}(b). Later on, MFR-1 stops its northward motion but expands and rotates into a higher altitude counterclockwisely (Figures \ref{f2}(b4)--(d4)), which may be caused by an interaction with the wrapping magnetic fields. Simultaneously, the spiral arms become longer and more evident as seen in the EUVI 304 \AA\ images (Figure \ref{f2}(a4)). When MFR-1 is erupting, expanding, swirling or rotating (i.e., untwisting), and disintegrating, the materials are injected into the upper atmosphere (Figures \ref{f2}(a5)--(d5) and (a6)--(d6)). The untwisting motion of MFR-1 lasts for about 20 minutes. At a late stage, there are some materials falling down along the helical trajectories (see the accompanying animation of Figure \ref{f2}).

\subsubsection{Kinematic motions}

In order to study the kinematic motions of MFR-1 from two viewpoints, we cut three slices (S1, S2, and S3 as indicated in Figures \ref{f2}(a4) and (b4)) in AIA and EUVI images to track the temporal evolution of MFR-1. We also utilize the Fourier local correlation tracking (FLCT) method \citep[]{Welsch2004,Fisher2008} to study the swirling or untwisting motion of MFR-1 using AIA images.

\begin{figure}[ht]
\centering
\includegraphics[width=0.9\textwidth]{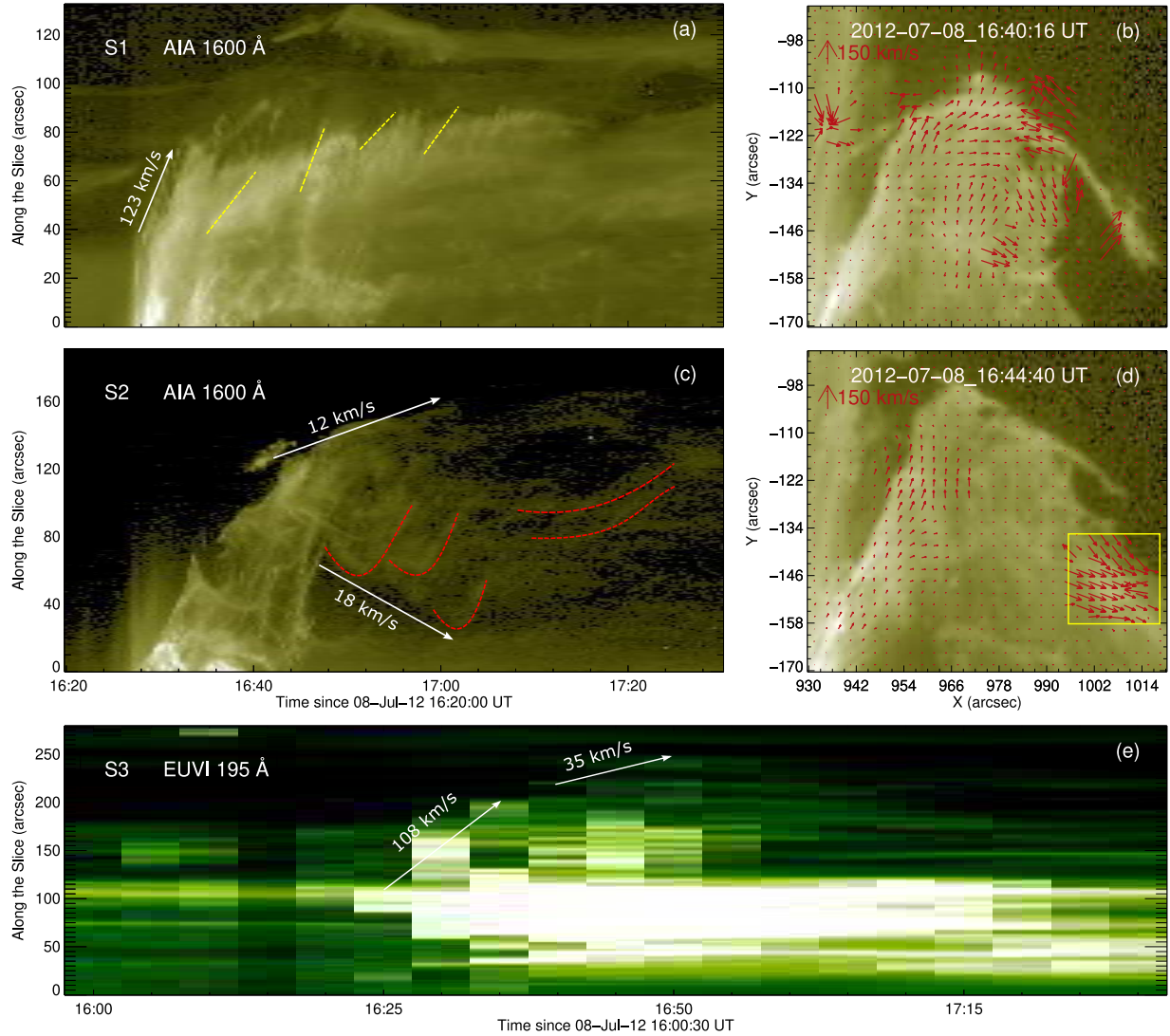}
\\[0mm]
\caption{Kinematic motions of MFR-1. Panels (a) and (c) show the time-slice diagrams of AIA 1600 \AA\ intensity along S1 and S2 (marked in Figure \ref{f2}(b4)), respectively. The yellow dashed lines in panel (a) denote the intermittent ejections of MFR-1 and the red dashed curves in panel (c) mark the swaying patterns or untwisting motions of MFR-1. Panels (b) and (d) display the FLCT maps of AIA 1600 \AA\ at two times with the red arrows mainly indicating the swirling motions and material ejections (particularly in the yellow box region) of MFR-1. Panel (e) gives the time-slice diagram of EUVI 195 \AA\ intensity along S3 (marked in Figure \ref{f2}(a4)). The white arrows in panels (a), (c), and (e) indicate eruption or expansion speeds of MFR-1.
}
\label{f3}
\end{figure}

Figure \ref{f3}(a) shows the time-slice map along S1 that is located at a lower part of MFR-1 from a side view. It is seen that MFR-1 has a movement towards north with a speed of about 123 km s$^{-1}$ at $\sim$16:28 UT, which indicates that MFR-1 is erupting. After that, there are materials intermittently ejected into the atmosphere from the northern footpoint of MFR-1 (see the yellow dashed lines). At the northernmost part of MFR-1, the materials show some swirling motions clockwisely, which can be clearly seen from the FLCT map in Figure \ref{f3}(b). The swirling speeds are from a few tens to more than one hundred km s$^{-1}$ with an average of about 55 km s$^{-1}$ at $\sim$16:40 UT. At $\sim$16:44 UT, some materials are injected into the upper atmosphere with a typical speed of $\sim$100 km s$^{-1}$, which can be seen from the FLCT map in Figure \ref{f3}(d) (marked by the yellow box). Besides the swirling motions at the northernmost part of MFR-1, some similar but counterclockwise rotations can be seen in the main body of MFR-1 afterwards. From the time-slice map along S2 in Figure \ref{f3}(c), one can see that there displays a swaying pattern (marked by some red dashed curves), indicating that MFR-1 is rotating along its axis. One can also see that the width of MFR-1 increases with a speed of 12--18 km s$^{-1}$ (see the two white arrows), which demonstrates that MFR-1 is expanding. The swirling, rotating, and expanding motions suggest that MFR-1 is untwisting. From the disk view of EUVI (Figure \ref{f3}(e)), we can see the eruption and expansion of MFR-1 as well. Starting from $\sim$16:25 UT, MFR-1 undergoes a rapid motion with a speed of about 108 km s$^{-1}$ towards north. At a later time ($\sim$16:40 UT), one can notice a slow motion with a speed of about 35 km s$^{-1}$ instead. Note that these speeds are just in the plane of sky and the real speeds should be somewhat larger due to the projection effect. The rapid motion at an earlier time is supposed to correspond to the eruption stage of MFR-1, and the slow motion after that basically corresponds to its expansion and untwisting stage. The fast-eruption stage lasts for about 10 minutes and the slow-untwisting stage lasts for a longer time (about 20 minutes).

\subsubsection{Thermal Property}

\begin{figure}[ht]
\centering
\includegraphics[width=0.9\textwidth]{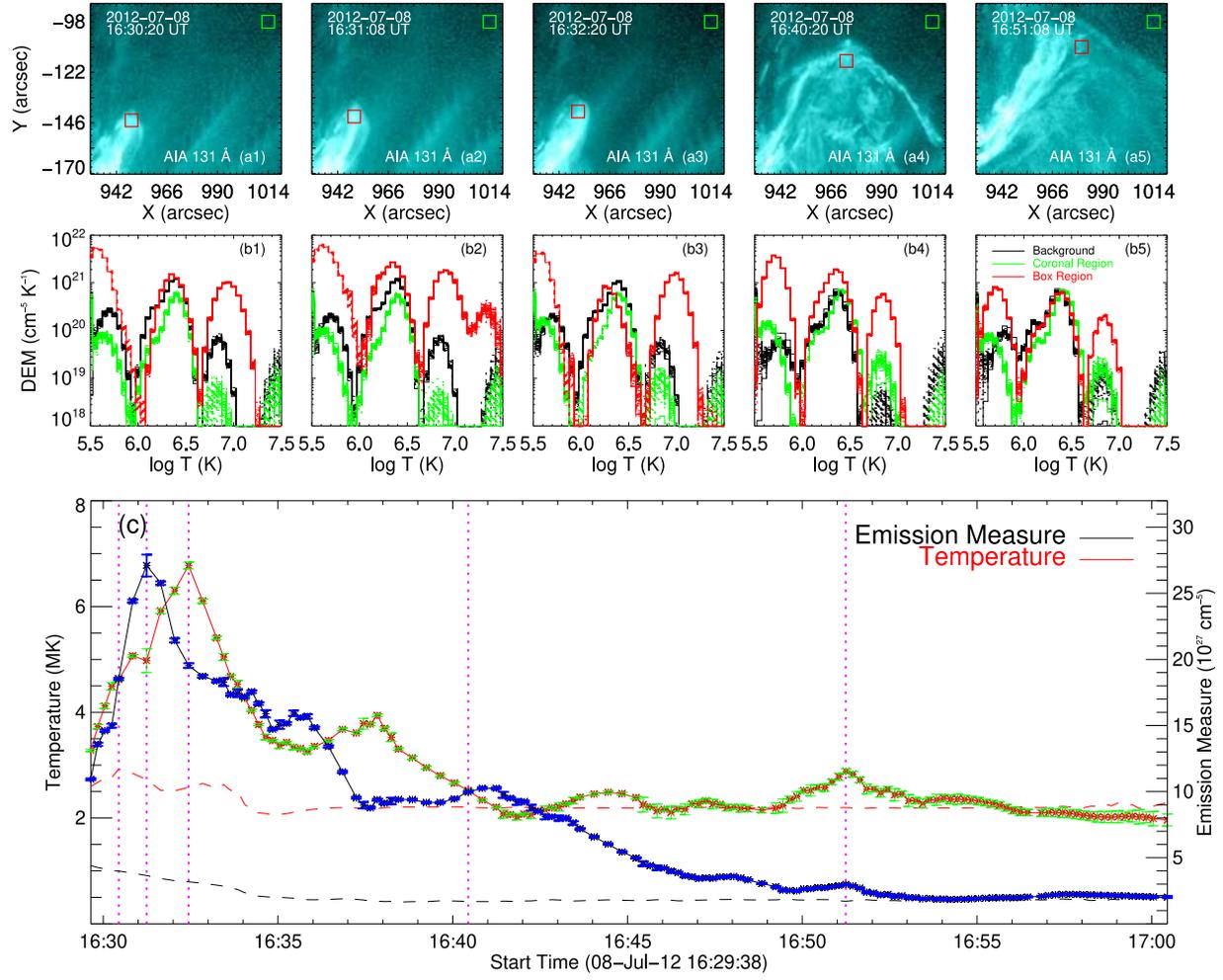}
\\[0mm]
\caption{DEM analysis of MFR-1. Panels (a1)--(a5) show the evolution of MFR-1 at AIA 131 \AA\ from $\sim$16:30 UT to $\sim$16:51 UT on 2012 July 8. The red box encloses some tracked materials from MFR-1 and the green box denotes a quiet coronal region nearby for comparison. Panels (b1)--(b5) plot the DEM distributions (red and green curves) for the red and green box regions at five times corresponding to the above AIA 131 \AA\ images. Note that the black curves also represent the DEM distributions from the red box region but before the eruption ($\sim$25 minutes earlier). The uncertainties of DEM are derived from 100 Monte Carlo simulations. In panel (c), the solid curves represent the temporal evolutions of EM and EM-weighted temperature for the red box region, with the error bars showing the uncertainties. Note that here EM is calculated by integrating DEM over a temperature range of log $T=$ 5.55--7.30. The dashed curves also represent the EM and temperature for the red box region but before the eruption. The five vertical magenta lines denote the five times of AIA 131 \AA\ images as shown in panel (a).
}
\label{f4}
\end{figure}

To investigate the thermal property of MFR-1 during its eruption and untwisting processes, we try to track some ejected materials visually and plot the DEM distributions together with EM and temperature evolutions in Figure \ref{f4}. From the AIA 131 \AA\ images (top panels), we can see that the tracked materials (enclosed by the red box) move northwards during $\sim$16:30--16:35 UT (Figures \ref{f4}(a1)--(a3)). When MFR-1 is obviously expanding and untwisting (after $\sim$16:40 UT, Figure \ref{f4}(a4)), these materials are fragmented and finally hard to be tracked (see Figure \ref{f4}(a5)). The DEM distributions for the red box region are shown in the middle panels of Figure \ref{f4} (see the red curves). For comparison, we also plot the DEM distributions for the red box region but before the eruption ($\sim$25 minutes earlier, black curves) serving as a background or reference. Moreover, we give the DEM results (green curves) for a quiet coronal region nearby (marked by the green box) for another reference, which are actually similar to the background curves. It can be seen that, relative to the background or quiet coronal region, the MFR-1 plasmas mainly show DEM enhancements peaking at $\sim$0.4--0.5 MK (log $T\approx\,$ 5.6--5.7) and $\sim$7--8 MK (log $T\approx\,$ 6.8--6.9) throughout the evolution. Note that these ejected plasmas also exhibit a much hotter component at some particular times, say, a component peaking at $\sim$20 MK (log $T\approx\,7.3$) around 16:31 UT (see Figure \ref{f4}(b2)). These results indicate that the plasmas of MFR-1 consist of a cold component and a hot component in principle and that there is some localized heating during its dynamic evolution. The evolutions of EM and EM-weighted temperature from the red box region are plotted in the bottom panel of Figure \ref{f4} (see the solid curves with error bars). One can see that both the EM and temperature have an increase followed by a decrease and finally return to the background level (dashed curves). This also demonstrates that the MFR-1 plasmas are heated during the evolution, especially in the early period. Note that the temperature reaches its maximum ($\sim$7 MK) a little bit later than EM, which might suggest that the plasmas are heated locally or in situ.

\begin{figure}[ht]
\centering
\includegraphics[width=0.9\textwidth]{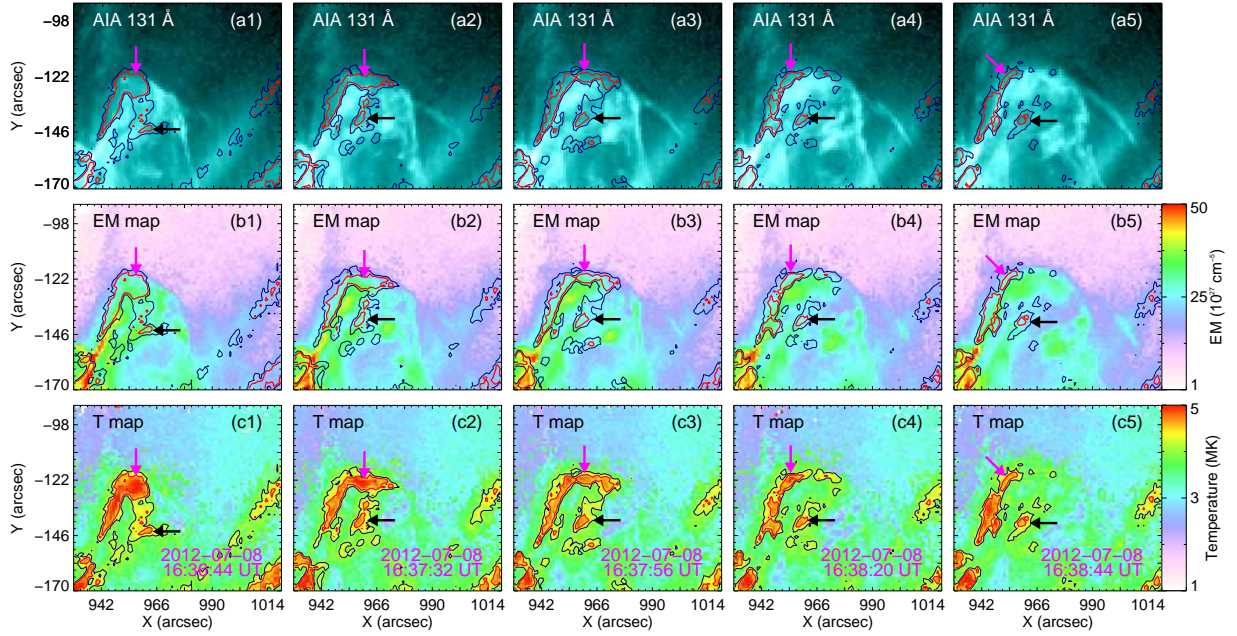}
\\[0mm]
\caption{AIA 131 \AA\ images (top row), EM maps (middle row), and EM-weighted temperature maps (bottom row) at five times for the untwisting process of MFR-1. The temperature contours at 4.0 (red) and 3.5 (navy) MK are overplotted on all of the maps. The magenta and black arrows indicate the hot regions at the north edge and in the center of MFR-1, respectively.
}
\label{f5}
\end{figure}

In addition, we show the spatial maps of EM and EM-weighted temperature during the untwisting process of MFR-1 in Figure \ref{f5}. From $\sim$16:35--16:40 UT, MFR-1 stops its northward motion but exhibits some swirling motions, as seen from the AIA 131 \AA\ images in the top panels of Figure \ref{f5}. The EM maps in the middle panels show similar structures to that in the AIA 131 \AA\ images. This can be expected since both the EM and EUV emissions are closely related to the electron density. The temperature maps overlaid with contours at 4.0 and 3.5 MK are given in the bottom panels of Figure \ref{f5}. The temperature contours are also overplotted on the AIA 131 \AA\ and EM maps. It is interesting to see that hot emissions mainly show up at the north edge (denoted by the magenta arrow). Some hot emissions can also be found in the center of or within MFR-1 (see the black arrow). These hot plasmas may be heated when MFR-1 is interacting with the ambient magnetic fields, say, via magnetic reconnection. Magnetic reconnection could also occur within MFR-1 and heats the plasmas therein. Another possibility is that the plasmas are compressed and thus heated when they are swirling or interacting with the ambient magnetic structures. 

\subsection{The MFR-2 on 2012 July 9}
\subsubsection{Morphological evolution}

\begin{figure}[ht]
\centering
\includegraphics[width=0.55\textwidth]{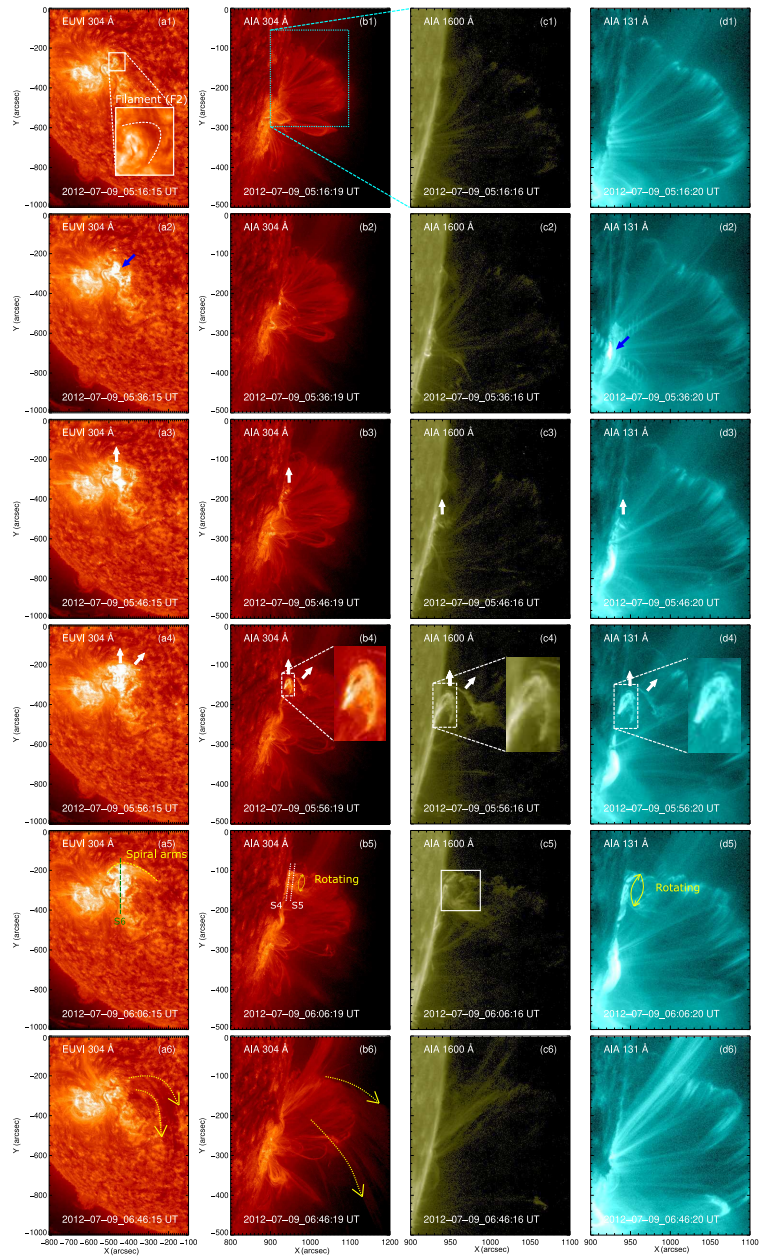}
\\[0mm]
\caption{Snapshots of the onset, development, and disintegration of the second event at EUVI 304 \AA, AIA 304, 1600, and 131 \AA\ (from left to right) from 05:16 UT to 06:46 UT on 2012 July 9. Note that the FOV of AIA 1600 and 131 \AA\ images is smaller than the one of EUVI and AIA 304 \AA\ images, which is denoted in panel (b1). In panel (a1), the filament (F2) is enlarged and indicated by a white dashed curve. The blue arrows in panels (a2) and (d2) mark the associated flare. The white arrows in panels (a3)--(d3) and also in panels (a4)--(d4) indicate the eruption and expansion of MFR-2. MFR-2 is also enlarged in panels (b4)--(d4). The yellow dashed curve in panel (a5) refer to some magnetic structures with a spiral arm. The three slices S4, S5, and S6, indicated in panels (a5) and (b5), are used to study the temporal evolution of MFR-2 as shown in Figures \ref{f7}(a), (c), and (e), respectively. The white box in panel (c5) encloses the region that is used to show the swirling motions and material ejections of MFR-2 in Figures \ref{f7}(b) and (d).The yellow solid arrows in panels (b5) and (d5) indicate the rotation of MFR-2 and the yellow dashed arrows in panels (a6) and (b6) mark the ejection of the materials from MFR-2.
The images with the same wavelength selection are available as an animation which presents a temporal evolution of MFR-2, including the onset, development, and disintegration processes, at EUVI 304 \AA\ (with a cadence of 10 min), AIA 304 (a cadence of 12 s), 1600 (a cadence of 24 s), and 131 \AA\ (a cadence of 12 s) (from left to right in the animation) from 05:06 UT to 06:56 UT on 2012 July 9. The real-time duration of this animation is about 18 s.
}
\label{f6}
\end{figure}

The second eruption event took place on 2012 July 9, about 12 hours later than the first event. Figure \ref{f6} shows some snapshots for its onset, development, and disintegration at EUVI 304 \AA\ as well as AIA 304, 1600, and 131 \AA\ from 05:16 UT to 06:46 UT. Similar to the first event, a reversed-C shape filament (F2) pre-exists in the active region, which is supposed to be reformed after F1 erupts. At $\sim$05:25 UT, F2 starts to erupt and brighten up. It is related to a C6.0 flare, as denoted by the blue arrow in Figures \ref{f6}(a2) and (d2). After its eruption, F2 exhibits a movement towards the north (Figures \ref{f6}(a3)--(d3)). During this process, F2 begins to expand and displays a clear helical structure or a $\gamma$-like shape, i.e., MFR-2 (see Figures \ref{f6}(b4)--(d4)). A few minutes later, we can see a cluster of spiral arms (outlined by the dashed curve in Figure \ref{f6}(a5)) showing up in the face view of EUVI images. From the side view of AIA images, it is seen that MFR-2 stops its northward motion then and exhibits a tornado-like structure that swirls or rotates clockwisely (see the accompanying animation of Figure \ref{f6} and Figure \ref{f7}(b)). This suggests that MFR-2 is untwisting. In the following, MFR-2 continues its clockwise rotation or untwisting motion with some of its plasmas ejected to the upper atmosphere (indicated by the yellow arrows in Figures \ref{f6}(a6) and (b6)). The untwisting motion of MFR-2 lasts for about 20 minutes. At a late stage, some materials fall down along the magnetic structures (see the accompanying animation). It should be pointed out that all of these dynamic behaviors are very similar to the first eruption event except that MFR-2 does not change its rotating direction from clockwise to counterclockwise before and after the plasma ejection.

\subsubsection{Kinematic motions}

\begin{figure}[ht]
\centering
\includegraphics[width=0.9\textwidth]{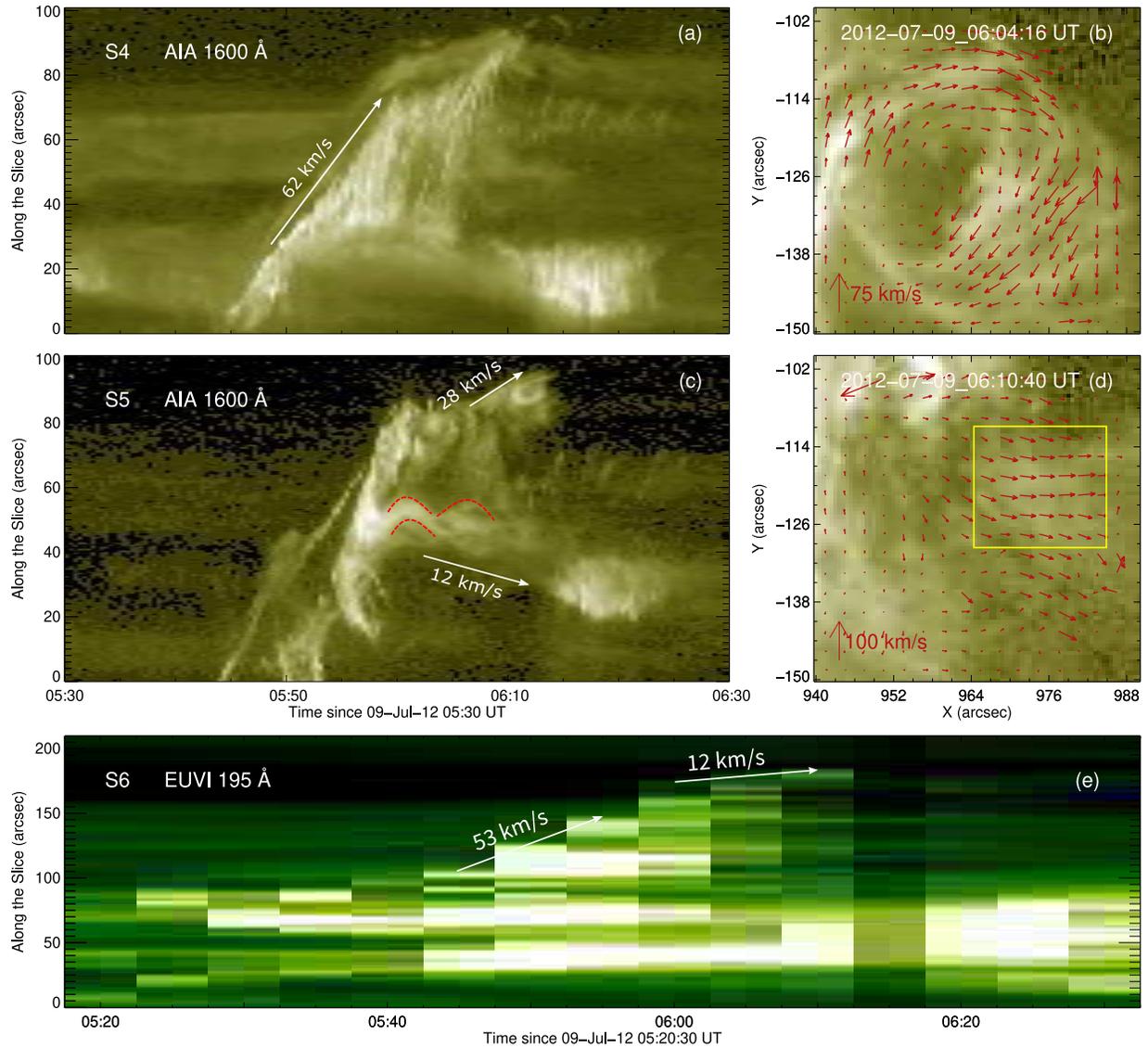}
\\[0mm]
\caption{Kinematic motions of MFR-2. Panels (a) and (c) show the time-slice diagrams of AIA 1600 \AA\ intensity along S4 and S5 (marked in Figure \ref{f6}(b5)), respectively. The red dashed curves in panel (c) mark the swaying patterns or untwisting motions of MFR-2. Panels (b) and (d) display the FLCT maps of AIA 1600 \AA\ at two times with the red arrows mainly indicating the swirling motions and material ejections (particularly in the yellow box region) of MFR-2. Panel (e) gives the time-slice diagram of EUVI 195 \AA\ intensity along S6 (marked in Figure \ref{f6}(a5)). The white arrows in panels (a), (c), and (e) indicate eruption or expansion speeds of MFR-2.
}
\label{f7}
\end{figure}

Similarly, we cut three slices (S4, S5, and S6 as indicated in Figures \ref{f6}(a5) and (b5)) in AIA  and EUVI images to study the kinematic motions of MFR-2 from two perspectives. We also utilize the FLCT method to study the swirling or untwisting motion of MFR-2 by AIA images.
Figure \ref{f7}(a) shows the time-slice map along S4 that is located at a bottom part of the tornado-like structure of MFR-2 as shown in images from a side view. We can see that MFR-2 erupts towards the north with a speed of about 62 km s$^{-1}$ at $\sim$05:44 UT. Later, it stops its northward motion and mainly shows a swirling motion as clearly seen from the FLCT map in Figure \ref{f7}(b). This may be caused by an interaction with the surrounding magnetic fields, similarly to the case of MFR-1. The swirling motion of MFR-2 is clockwise with an average speed of about 40 km s$^{-1}$ at 06:04 UT. From the time-slice map along S5 as shown in Figure \ref{f7}(c), it is seen that after $\sim$06:00 UT, MFR-2 continues rotating clockwisely but shows an expansion with a speed of about 12--28 km s$^{-1}$. At $\sim$06:10 UT, the plasmas in MFR-2 are ejected to the upper atmosphere with a typical speed of a few tens of km s$^{-1}$ (see the yellow box in Figure \ref{f7}(d)). The eruption and expansion (untwisting) can also be reflected in the disk view of EUVI images. From the time-slice map along S6 in Figure \ref{f7}(e), one can see a fast eruption of MFR-2 with a speed of about 53 km s$^{-1}$ before $\sim$06:00 UT. After that the speed changes to $\sim$12 km s$^{-1}$ when MFR-2 untwists and ejects plasmas to a higher altitude. Note that all of these speeds of MFR-2 are smaller than the corresponding ones of MFR-1. Both the fast-eruption stage and the slow-expansion stage of MFR-2 last for about 15 minutes. After $\sim$06:15 UT, MFR-2 disintegrates and some ejected materials fall down along the magnetic structures.

\subsubsection{Thermal Property}

\begin{figure}[ht]
\centering
\includegraphics[width=0.9\textwidth]{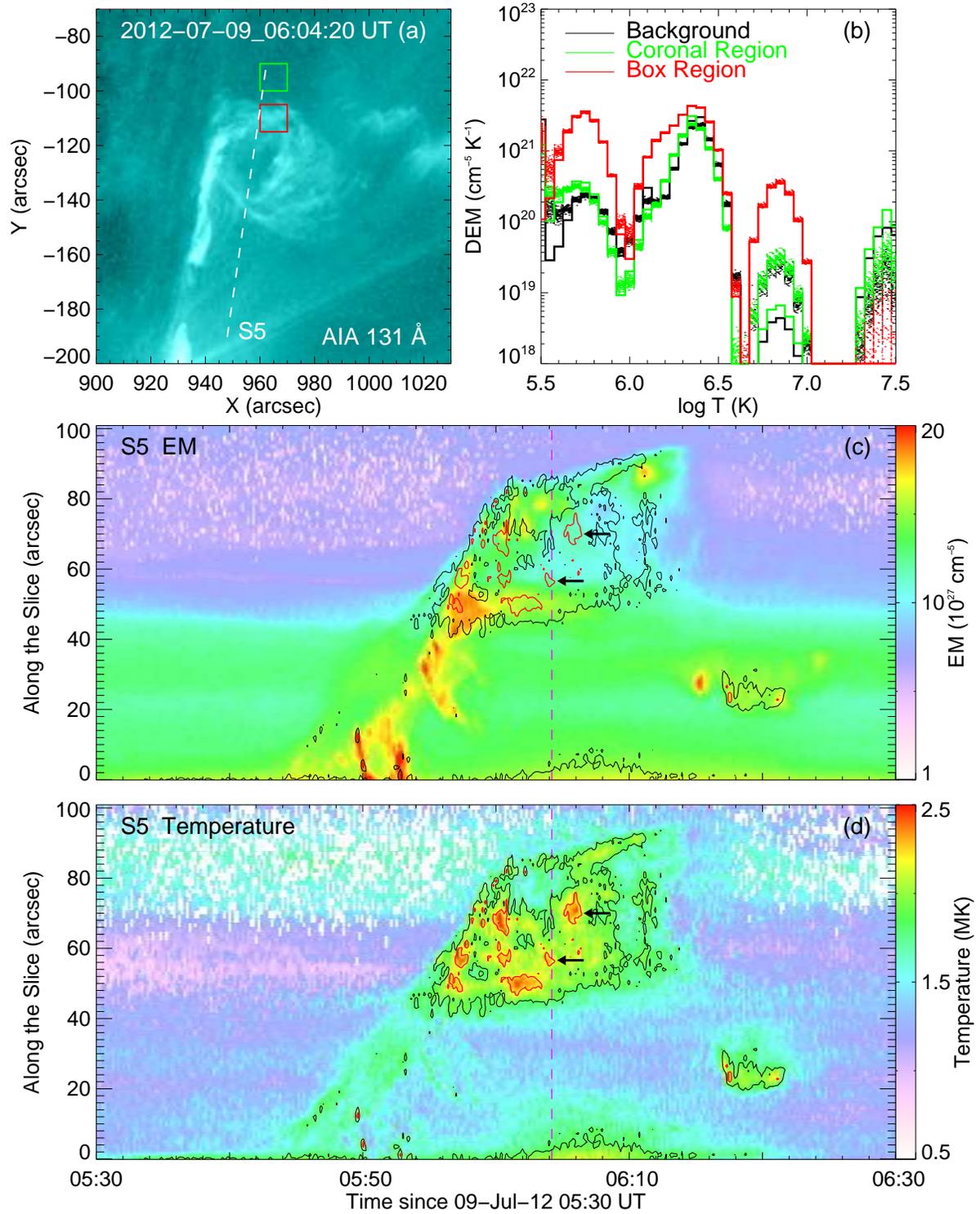}
\\[0mm]
\caption{DEM analysis of MFR-2. Panel (a) shows the AIA 131 \AA\ image of MFR-2 at 06:04:20 UT. The red box region is selected to study the DEM distribution of MFR-2 and the green box region nearby, i.e., a quiet coronal region, serves as a reference. Panel (b) plots the DEM distributions (red and green curves) for the red and green box regions marked in panel (a). The black curve also represents the DEM distribution from the red box region but before the eruption ($\sim$30 minutes earlier). The uncertainties of DEM are derived from 100 Monte Carlo simulations. Panels (c) and (d) show the time-slice diagrams of EM and temperature along S5 (also denoted in panel (a)), respectively. The temperature contours at 2.0 (red) and 1.5 (black) MK are overplotted in both diagrams. The black arrows indicate the high temperature feature in the center of MFR-2. The vertical magenta line marks the time for the AIA 131 \AA\ image in panel (a).
}
\label{f8}
\end{figure}

\begin{figure}[ht]
\centering
\includegraphics[width=0.9\textwidth]{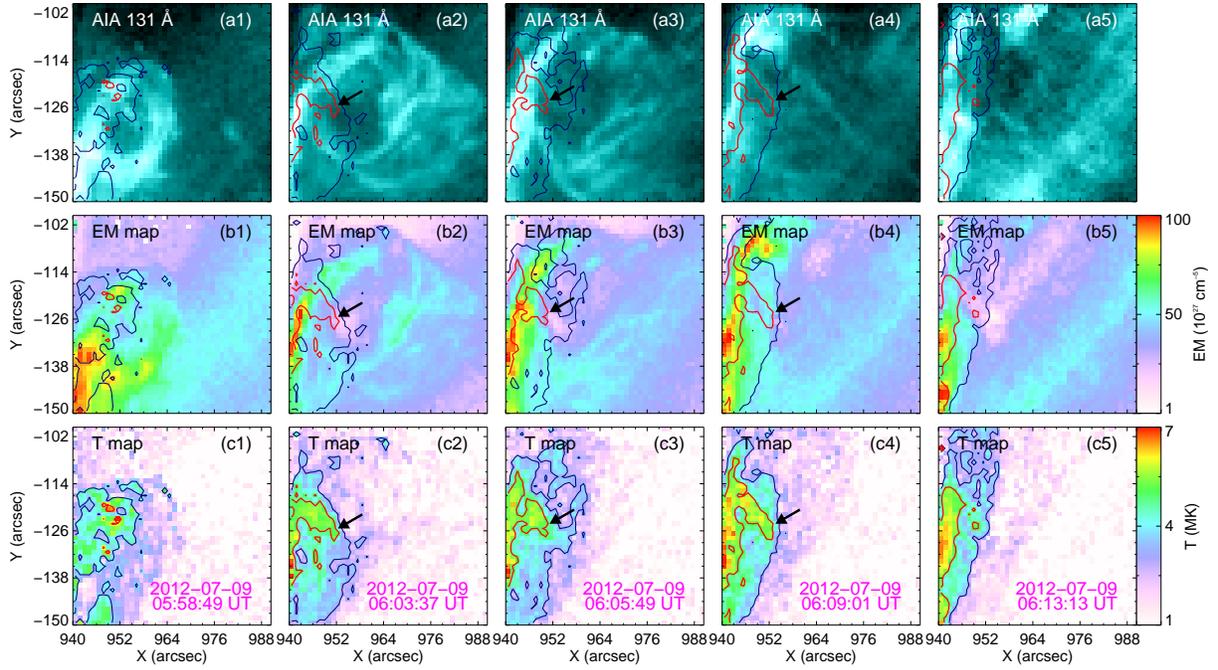}
\\[0mm]
\caption{AIA 131 \AA\ images (top row), EM maps (middle row), and EM-weighted temperature maps (bottom row) at five times for the untwisting process of MFR-2. The temperature contours at 3.5 (red) and 2.0 (navy) MK are overplotted on all of the maps. The black arrows denote the hot regions in the center of MFR-2.
}
\label{f9}
\end{figure}

To study the thermal property of MFR-2 particularly during its untwisting process, we plot the DEM distribution as well as EM and EM-weighted temperature maps in Figures \ref{f8} and \ref{f9}. Firstly, we select a region (marked by the red box in Figure \ref{f8}(a)) at the north edge of MFR-2 at $\sim$06:04 UT and show its DEM distribution (red curve) in Figure \ref{f8}(b). For comparison, we also plot the DEM curve (black) for the red box but before the eruption ($\sim$30 minutes earlier) serving as a background or reference. Moreover, we provide the DEM curve (green) from a quiet coronal region nearby (denoted by a green box in Figure \ref{f8}(a)) for a second reference. One can see that compared with the background and quiet coronal region, the DEM distribution enclosed by the red box has two main components peaking at $\sim$0.6 MK (log $T\approx\,$5.8) and $\sim$7 MK (log $T\approx\,$6.8), i.e., a cold component plus a hot component. This is also similar to the result of MFR-1, though the temperature of the hot component is slightly lower than that of MFR-1. Secondly, we show the evolutions of EM and temperature along S5 in Figures \ref{f8}(c) and (d), respectively. It is seen that the EM diagram basically exhibits the expansion (or untwisting) motion of MFR-2, as similarly shown in the AIA 1600 \AA\ images in Figure \ref{f7}(c). The interesting thing is that the temperature feature does not quite match the EM feature, say, a relatively high temperature (\textgreater2.0 MK) appears fragmentarily in the center of MFR-2 where EM is not so high (see the red contours overlaid on both the temperature and EM diagrams). We further plot the spatial maps of EM and temperature in Figure \ref{f9}, both of which are overlaid by temperature contours with levels of 3.5 and 2.0 MK. One can see that during the untwisting process, a relatively high temperature shows up mainly in the bottom part of MFR-2. In particular, the high temperature region extends to the center of MFR-2 (marked by the black arrow) where EM is relatively low. All these results suggest that some heating processes are working within MFR-2, probably due to internal magnetic reconnection or plasma compression.

\section{Summary and Discussions}
\label{sum}

In this paper, we study two homologous MFRs in NOAA 11515 using multi-waveband and dual-perspective imaging observations from SDO/AIA and STEREO-A/EUVI. The morphological evolution, kinematic motions, and thermal property of the two MFRs are analyzed. The observational features of the two MFRs are summarized in Table \ref{tab1}. Our main findings are as follows. (1) Both of the MFRs show up in multi-wavelength passbands and their DEM distributions mainly consist of a cold component peaking at $\sim$0.4--0.6 MK and a hot component peaking at $\sim$7--8 MK. (2) The two MFRs exhibit erupting, expanding, and untwisting motions and their evolution can be divided into two stages, a fast-eruption stage with speeds of a few tens to hundreds of km s$^{-1}$ and a slow-expansion (or untwisting) stage with speeds of several tens of km s$^{-1}$. (3) During the two-stage evolution, hot plasmas show up at the edge and in the center of the two MFRs, indicating that some local heatings take place there via magnetic reconnection and/or plasma compression.

\begin{table}[htb]
\begin{center}
\small
\caption{Summary of the observational features of the two homologous MFRs}
\label{tab1}
\begin{tabular}{lcc}
\hline
\hline
MFRs  & MFR-1 & MFR-2 \\
\hline
Observing Dates & 2012-07-08 & 2012-07-09 \\
Associated Filaments & F1 (pre-existing) & F2 (pre-existing) \\
Related Flares & M6.9 & C6.0 \\
Erupting Motions$^*$ & northward (asymmetric) & northward (asymmetric) \\
                                   & 123 km s$^{-1}$ (AIA) & 62 km s$^{-1}$ (AIA) \\
                                   & 108 km s$^{-1}$ (EUVI) & 53 km s$^{-1}$ (EUVI) \\
Expanding Motions$^*$ &12--18 km s$^{-1}$ (AIA) & 12--28 km s$^{-1}$ (AIA) \\
                                      & 35 km s$^{-1}$ (EUVI) & 12 km s$^{-1}$ (EUVI) \\
Swirling/Untwisting & $\sim$55 km s$^{-1}$ (AIA) & $\sim$40 km s$^{-1}$ (AIA) \\
Motions$^*$         & clockwise first & clockwise first \\
                    & then counterclockwise & then still clockwise \\
Plasma Emission & multi-temperature & multi-temperature \\
                             & $\sim$0.4--0.5 MK \& $\sim$7--8 MK & $\sim$0.6 MK \& $\sim$7 MK \\
High Temperature & at the northern edge & at the bottom edge \\
Features                & and in the center & and in the center \\
\hline
Additional Comments & \multicolumn{2}{c}{in a similar magnetic environment before eruption} \\
                                & \multicolumn{2}{c}{with plasmas ejected during eruption and falling back at late stage} \\
\hline
\hline
\end{tabular}
\end{center}
$^*$ All the speeds of motions are measured in the plane of sky.
\end{table}

\subsection{General Remarks on the Observational Features of the Two MFRs}

The two homologous MFRs show some similarities and also differences in the observational features as seen from Table \ref{tab1}. More specifically, both MFRs are associated with filaments and flares and show erupting, expanding, and untwisting motions, though the flare magnitudes and the speeds of motions are somewhat different. The morphological evolutions of the two MFRs are almost identical as seen from the EUVI viewpoint. In addition, the DEMs of both MFRs consist of a cold component plus a hot component with the high temperature features mainly appearing in the interface region between the MFRs (their northern or bottom edge) and the ambient magnetic structures as well as in the center of the MFRs. We ascribe these similarities to a similar magnetic environment in which the two MFRs are rooted, as manifested by the similar reversed-C shape filaments and especially the surrounding magnetic fields with a large spiral arm before the eruptions. On the other hand, some noticeable differences, including the morphology of the two MFRs and the directions of rotation before and after the plasma ejection, can be clearly seen in the AIA images. These differences can only be distinguished by multi-viewpoint observations that are necessary to fully understand the eruption process of MFRs.

\subsection{Roles of the Ambient Magnetic Fields in MFR Eruptions}

In both of the events, the MFRs move northward (i.e., asymmetric eruptions) at an early stage. During this process, the MFRs are supposed to interact with the ambient magnetic fields especially on the north side based on the following signatures. (1) Before the eruptions, some large-scale helical structures (as background magnetic fields) are clearly seen in the EUVI 195 \AA\ images. (2) During the eruptions, the MFRs stop moving towards the north probably constrained by the ambient magnetic fields. (3) MFR-1 changes its rotating direction from clockwise to counterclockwise before and after the plasmas are ejected to the upper atmosphere along the helical trajectories. Note that MFR-2 does not change its rotating direction during the eruption. (4) The two MFRs show some heatings at the interface region, i.e., at the north or bottom edge, which are probably caused by magnetic reconnection between the MFRs and ambient magnetic fields during the eruptions. In other words, the ambient magnetic fields play an important role in MFR eruptions, say, constraining the MFRs, changing the directions of MFR motions, increasing the twist of MFRs, heating MFRs, and transferring the twist and materials from MFRs to the ambient magnetic loops, as have been reported by previous studies \citep[e.g.,][]{Ji2003,Liu2007,Cohen2010,Bi2013,Yanglh2019,Yan2020b,Yan2020a}. It should be noted that in the two events under study, the strength of interaction between the MFRs and ambient magnetic fields is supposed to be different, which probably depends on the magnetic field strength (or magnetic topology) and also erupting speeds of the MFRs and so on. The interaction of MFR-1 with the ambient magnetic structures could be more intense than the one of MFR-2, as MFR-1 has a larger erupting speed than MFR-2. A stronger constraining force from the ambient magnetic fields might also play a role in MFR-1, while in MFR-2 the constraining force could become weaker or the magnetic topology has somewhat changed after MFR-1 erupts. Therefore, MFR-1 changes its rotating direction during the eruption while MFR-2 does not.

\subsection{Eruption of the Two MFRs}

\begin{figure}[ht]
\centering
\includegraphics[width=\textwidth]{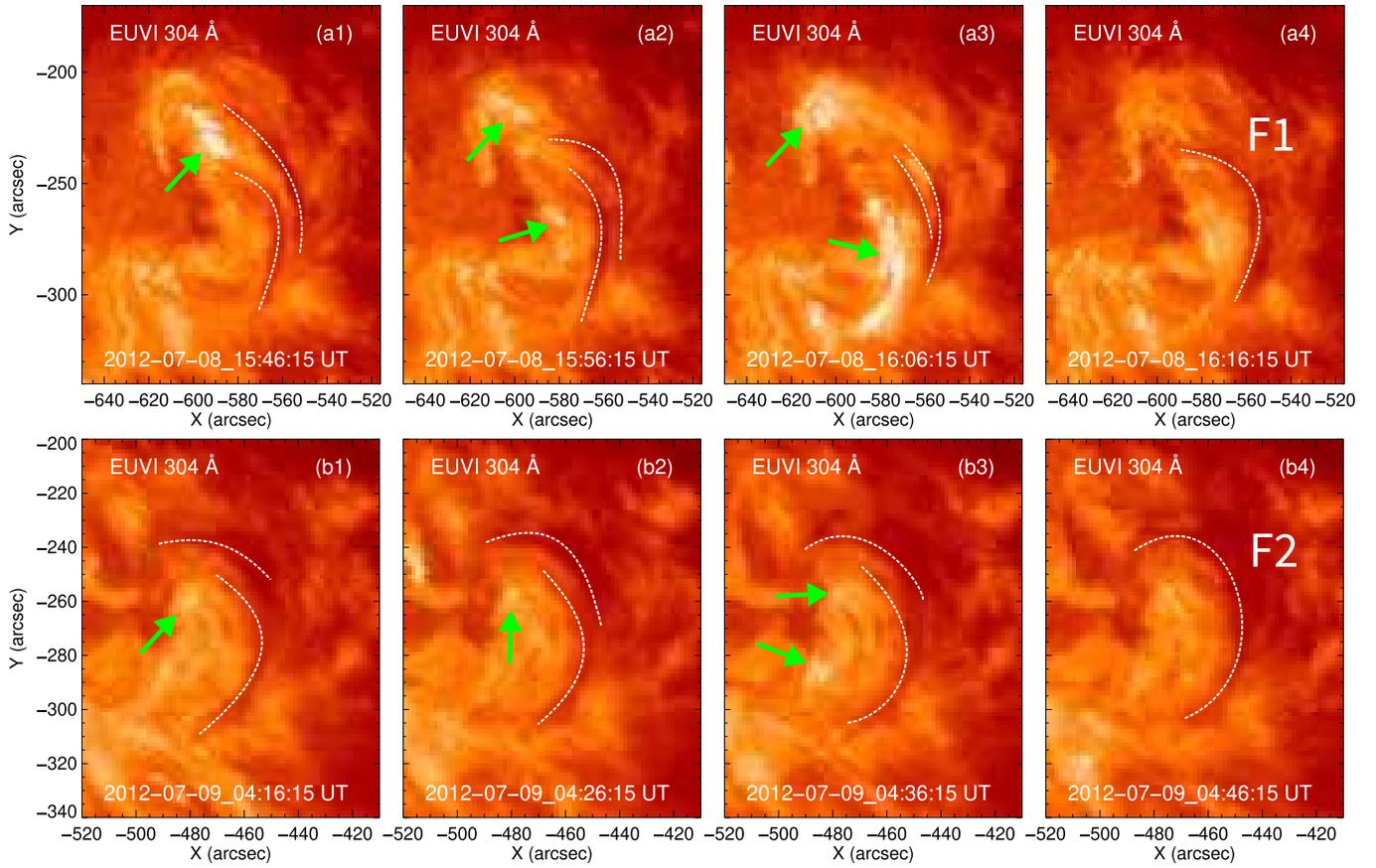}
\\[0mm]
\caption{Pre-flare signatures of tether-cutting reconnection for the two events. The white dashed curves in panels (a1)--(a3) and (b1)--(b3) denote separate curved structures with time evolution and the green arrows mark the brightenings around them. The white dashed curves in panels (a4) and (b4) indicate the reversed-C shape filaments of F1 and F2 just before the eruptions. 
}
\label{f10}
\end{figure}

There are some interesting topics or questions related to the eruption of MFRs such as trigger mechanisms and eruption characteristics. Here we discuss some of them for the two homologous MFRs in the present study. 
Firstly, what lead to the onset of the two MFRs toward eruption, i.e., the trigger mechanisms? It has been widely accepted that magnetic reconnections such as tether-cutting and breakout types as well as ideal MHD processes including kink or torus instabilities can trigger solar eruptions \citep[e.g.,][]{Moore2001,Torok2005,Kliem2006,Aulanier2010,Shen2012breakout,Zuccarello2014}. In our two events studied here, we speculate that tether-cutting reconnection and kink instability likely play roles in the MFR eruptions according to the following observational features. Tens of minutes before the eruption, both filaments display separate curved structures and some brightenings appear around simultaneously (as shown in Figures \ref{f10}(a1)--(a3) and (b1)--(b3)). We then clearly see a reversed-C shape filament, namely F1 or F2, just before the eruption (Figures \ref{f10}(a4) and (b4)). This may suggest that tether-cutting reconnection takes place, which can help form the filaments as well as push the filaments into an unstable state (say, via increasing the twist). When the filaments start to erupt, they show a $\gamma$-like shape or a writhed structure, which may indicate that a kink instability is happening. Note that here we cannot rule out the other trigger mechanisms. Secondly, what is the role of the reversed-C shape filaments in the eruption? As mentioned above, when the reversed-C shape filaments are formed via tether-cutting reconnection, their twist can increase during the process. Once the twist increases to a certain value, the filament will writhe by the conversion of twist and writhe \citep[e.g.,][]{Kliem2010}. As a result, we see the $\gamma$-like structure in the eruption as well as a rotation of the MFR afterwards caused by kink instability. The reversed-C shape of filaments also determines the initial clockwise direction of the MFR rotation. Thirdly, why do the MFRs show a fast-eruption stage followed by a slow-expansion stage? This could be explained in two aspects. On the one hand, both MFRs experience a kink instability and show an asymmetric eruption toward north. During this course, the MFR can be accelerated, at least in the initial time. As moving northward further, the MFRs are constrained by the ambient magnetic fields, which can change their kinematic motions, say, entering a relaxation (or an expansion) stage from the eruption phase. On the other hand, when the MFRs interact or reconnect with the ambient magnetic fields, their twist could increase further. When the twist of MFRs exceeds a certain threshold, the MFRs would become unstable \citep[e.g.,][]{HP1981,Baty2001,Torok2005,Williams2005,Srivastava2010} and an untwisting process could happen \citep[e.g.,][]{Alexander2006,Li2015}. Therefore, the MFRs show a fast-eruption stage first and then a slow expansion/untwisting stage during the eruption.

\subsection{Untwisting of the Two MFRs}

The two MFRs studied here exhibit prominent untwisting motions which could provide some information on their magnetic structures such as the twist. It is a consensus that MFRs consist of helical magnetic fields and the toroidal current dominates the twist which is related to the untwisting of MFRs. In practice, one could estimate the twist number of MFRs from their untwisting motions. Some previous studies have reported untwisting motions of MFRs and estimated their twist number using high resolution observations. For example, \cite{Yan2014b} detected counterclockwise untwisting motions of an active region filament and derived a total twist of at least 5$\pi$ by using the time slice method. \cite{Li2015} also estimated the total twist (about 4$\pi$) of an MFR from its untwisting motion. For the two MFRs in the present study, the twist is estimated to be at least one turn (i.e., 2$\pi$) from the time-slice diagrams (see the red dashed curves in Figures \ref{f3}(c) and \ref{f7}(c), each of which represents half a turn and a pair from the same location can give one turn). In addition, from the accompanying animations we could see the MFRs rotate more than one turn. Note that one turn is just the lower limit of the twist number for the two MFRs and the real twist number could be much greater than this, say, reaching a strongly kink-unstable threshold of 5$\pi$ \citep{Kliem2012}. It is a pity that we cannot obtain the twist number accurately using the time-slice method here since the MFR emissions are somewhat weak during the untwisting process especially at a higher height.

\subsection{Heating of the Two MFRs}

Through the DEM analysis, we find that the DEMs of both MFRs contain a hot component peaking at $\sim$7--8 MK and that the hot plasmas are mainly located in the interface region between the MFRs and the ambient magnetic fields and also in the center of MFRs. This result indicates some heating effects in the MFR plasmas during the eruption and untwisting processes. Considering that the twist of MFRs seems to increase at first (and is released later) when the MFRs interact with the ambient magnetic fields, we speculate that magnetic reconnection takes place between the MFRs and ambient magnetic fields that heats the plasmas at the north or bottom edge of MFRs \citep[e.g.,][]{Yan2013,Yanglh2019}. The plasmas in the center of MFRs could be heated by internal magnetic reconnection that occurs within the MFRs \citep[e.g.,][]{Galsgaard1997,Gibson2006,Gibson2008,Fermo2014,Yang2015,Mei2020ApJ}. There is another possibility that the MFR plasmas are compressed so as to be heated when the MFRs interact with the ambient magnetic fields and rotate around their axes. Note that in the two MFRs studied here, we think that the plasmas are more likely heated locally during but not before their eruptions based on the result that the temperature of some tracked plasmas reaches its maximum a little bit later than EM.

\acknowledgments
This work greatly benefits from the high quality data from SDO and STEREO. We acknowledge the use of DEM and FLCT codes. We thank Dr. Ying-na Su, Dr. Yang Guo, Dr. Zhi-xin Mei, and Dr. Jin-cheng Wang for their valuable discussions. We also thank the anonymous referee for the very constructive comments and suggestions that improve the manuscript. The authors are supported by NSFC under grants 11873095, 11733003, 11961131002, 11921003, and U1731241, and by the CAS Strategic Pioneer Program on Space Science under grants XDA15052200, XDA15320103, and XDA15320301. Y.L. is also supported by the CAS Pioneer Talents Program for Young Scientists.

\bibliography{homo}{}
\bibliographystyle{aasjournal}

\end{document}